\documentclass[doublecol]{epl2} 
\usepackage{amsmath}
\usepackage{epstopdf}
\usepackage{eurosym}

\title{Exponential and power laws in public procurement markets}

\author{Ladislav Kristoufek\inst{1,2},Jiri Skuhrovec\inst{1}}
\institute{                    
  \inst{1} Institute of Economic Studies, Charles University, Opletalova 26, Prague, CZ-110 00, Czech Republic, EU\\
  \inst{2} Institute of Information Theory and Automation, Academy of Sciences of the Czech Republic, Pod Vodarenskou vezi 4, Prague, CZ-182 08, Czech Republic, EU
}


\abstract{For the first time ever, we analyze a unique public procurement database, which includes information about a number of bidders for a contract, a final price, an identification of a winner and an identification of a contracting authority for each of more than 40,000 public procurements in the Czech Republic between 2006 and 2011, focusing on the distributional properties of the variables of interest. We uncover several scaling laws -- the exponential law for the number of bidders, and the power laws for the total revenues and total spendings of the participating companies, which even follows the Zipf's law for the 100 most spending institutions. We propose an analogy between extensive and non-extensive systems in physics and the public procurement market situations. Through an entropy maximization, such {the} analogy yields some interesting results and policy implications with respect to the Maxwell-Boltzmann and Pareto distributions in the analyzed quantities.}

\begin{document}

\maketitle

{Analyzing} distributional properties of different phenomena in social and economic systems has become popular in recent years ranging from the historically most popular wealth and income distributions {\cite{Pareto1896,Mandelbrot1961,Slanina2004,Coelho2008,Fiaschi2012}} to productivity \cite{Aoyama2010}, city size {\cite{Benguigui2007,Cordoba2008,Levy2009,Giesen2010}}, firm's size \cite{Stanley1995}, growth \cite{Salinger1996} and bankruptcy \cite{Fujiwara2004}, internet \cite{Adamic2002}, financial returns and volatility \cite{Mantegna1995,Gabaix2003,Gabaix2006}, traded volume \cite{Souza2006} {and most recently, several social and economic phenomena have been analyzed with Internet-based measures \cite{Saavedra2011,Preis2012,Vespignani2009}}. {See \cite{Farmer2008,Gabaix2008,Lux2008} for recent reviews.} One of the topics absolutely untouched by such a statistical analysis is the public procurements market. Analysis of this market is absolutely crucial from economic, political and social point of view because huge sums of public money (collected from taxes) flow from state to private firms every year. According to OECD{\footnote{Organization for Economic Co-Operation and Development.}} \cite{OECD2011}, public procurement{s totaled to an average of} 17 \% of GDP{\footnote{Gross Domestic Product -- a total value of all final goods and services produced in a country during a time period.}} in the OECD member countries, making {the} government and state-owned enterprises the most significant buyer{s} in virtually every developed economy. Hence, the topic {a} has very high economic relevance, yet {the} related research has only been very sparse so far{, mainly} due to a low availability or quality of {the relevant} data. We have overcome this problem {to some extent} and {have} obtained {a} broad and reliable dataset. This paper takes a natural first step in its examination usually characteristic for statistical physics -- while studying data distributions and statistical properties, we obtain economically relevant findings and directions for further research.

We start with several intuitive definitions.  {A} \textit{public procurement} (PP) is a specific procedure {of} purchasing goods and services, which is mandatory for various public institutions -- municipalities, government bodies, state-owned enterprises, etc., jointly called \textit{contracting authorities}. During {the} PP procedure (\textit{a tender}), various companies place their \textit{bids} -- offers to provide goods requested by {the} contracting authority for {a specific} price. One of these bids is then chosen by the contracting authority, we call the company which placed the bid either {\textit{a supplier}} or {\textit{a winner}}. In this paper, we study {the} distributional properties of three important quantities in {the} PP -- {a} number of bidders, total revenues of {the} individual suppliers and total spendings of {the} contracting authorities.

We focus on two laws standardly observed across {scientific} disciplines -- {the} exponential and power laws. Let us define a cumulative distribution function, \textit{cdf}, as $F(x)=P(X\ge x)$. {The p}ower law is described as $F(x) \propto x^{-\alpha}$ with a power-law exponent $\alpha$ and is usually labeled as the Pareto law or distribution. Corresponding probability density function, $pdf$, is defined as $f(x)=\partial F(x)/\partial x \propto x^{-(\alpha+1)}$. {The e}xponential law is then {characterized by} $F(x) \propto \exp(-\beta x)$ with an exponent $\beta$ and is often labeled as the Maxwell-Boltzmann distribution with an inverse temperate $\beta$ and {the} corresponding $cdf$ of $f(x)\propto \exp(-\beta x)$. The Pareto distribution is connected to the extensively analyzed Zipf's law, which is a power law between a rank and some other variable important for the analyzed system. If $F_i$ is a magnitude of some variable and $r$ is a corresponding rank, then $F_i\propto r^{-\gamma}$ is the Zipf's law. The Zipf's law is usually considered only for a special case when $\gamma=1$. It turns out that the power-law exponent $\alpha$ and the Zipf's law exponent $\gamma$ are inverse, i.e. $\alpha=1/\gamma$ \cite{Adamic2002}. As will be shown later, these two distributions are particularly important from economic point of view because they are {the} entropy-maximizing distributions of extensive and non-extensive systems, which can be well connected to the PP market.

The basic dataset{\footnote{Full dataset is available upon request from the authors; for its examination, at least rough knowledge of the European procurement law is necessary.}} has been obtained using web crawlers as a complete image of the public database ISVZ{\footnote{Information System about Public Procurement, isvzus.cz.}}, which contains all the Czech public tenders above a threshold of {an} expected price of {$6*10^6$ CZK ($\approx$ \euro$240*10^3$ or \$317$*10^3$)} for construction services and {$2*10^6$ CZK ($\approx$ \euro$80*10^3$ or \$$106*10^3$)} for all other procured goods or services. The full dataset {underwent} both automated and manual validity checks, assuring {the} mostly proper identification of contracting authorities, winners but also validity of other data fields. The data has been cross-checked against the company registry and further enriched using other public databases. {The d}ataset covers over 40,000 tenders from the period between 6/2006 and 8/2011. Due to features like {an} inclusion of small tenders, {a} coverage of {a} nationwide set of various tenders and most importantly {a} good data quality (which is highly above the European standards {of publication in TED\footnote{Tenders Electronic Daily, the official database of tenders in the EU.}}), the robustness of {the} results is ensured. Additionally, since the dataset is of {an} almost unique quality in this field and since the examined procurements follow the standard EU directives{\footnote{As described by the EU directive 2004/18 on the procurement of public works, public supply and public services contracts \cite{eu2004}.}}, our results are relevant {at least} Europe-wide but due to similarities in various procurement regulations possibly also outside the EU -- including the {USA} and Japan.
 
Let us now focus on {the} results for {the} number of bidders, total winner revenues and total contracting authority spendings.

{The c}umulative distribution function for {the} number of bidders is shown in Fig. \ref{fig1}. Almost a perfect fit in a linear-log scale indicates that the $cdf$ of the number of bidders is very well described by the exponential distribution with $\beta \approx 0.27$, which is supported for both $cdf$ and $pdf$. The most probable {(the most frequent)} number of bidders is a single bidder and the probability decreases exponentially. Approximately 95\% of the public procurements have 10 or less bidders. However, there is no intuitive or even basic economic reason for such a distribution to occur. Later, we propose that such a distribution emerges in {extensive} systems with suitable constraints {related to this specific problem}.

Compared to the number of bidders for a specific contract, {the} total revenues and total spendings range widely \footnote{Note that we consider only subjects with total spendings or revenues with at least $2*10^6$ CZK -- a floor amount above which the procurement has to be publicly listed.}. For {the} total revenues, the sums range from $2*10^6$ CZK up to $4*10^{10}$ CZK, and for {the} total spendings, the sums range from $2*10^6$ CZK to $1.5*10^{11}$ CZK. As the power law is defined only for one of the tails (it cannot hold for the whole distribution), we analyze {the} potential power law for {the} values above one standard deviation for both {the} total revenues and total spendings.

In Fig.~\ref{fig2}, we show the $cdf$ and Zipf plot for {the} total revenues. Both the log-log specified charts imply {the} power-law scaling with $\alpha \approx 1.24$ again with a practically perfect fit. Scaling in the Zipf plot indicates that, at least for the top 100 companies (with the highest revenues), the total revenues can be very well described by the Zipf's law with $\gamma \approx 0.79$ and the distribution of revenues is hence not uniform. {The e}mergence of such a scaling law indicates that the process is governed by a complex dynamics and interactions between competing agents. Such {an} interpretation is further developed later in the text. Similar behavior is observed for {the} total spendings on {the} public procurement contracts. Fig.~\ref{fig3} uncovers that {the} total spendings actually follow the exact Zipf's law with $\alpha \approx \gamma \approx 1$, i.e. the company with the second highest spendings has half the amount of the most spending company, the third company spends one third of the highest spendings, etc. Such a precise power law distribution again indicates that the whole process is governed by a complex dynamics and interactions between participating companies. Note that the documented power law exponents for revenues and spendings are not markedly different from {the} power laws observed for {the} income and wealth distributions \cite{Gabaix2008}. Also, as the Zipf's exponent is higher for {the} total spendings than for {the} total revenues, we can state that {the} distribution of money related to {the} public procurement is less equal for the contracting authorities than for the competing firms, which is rather unexpected. This is also well documented {by} the Lorenz curve (not shown here) which uncovers that the top 10\% of the competing firms obtains around 80\% of {the} total public procurement money (and the top 1\% of {the} firms still gets around 45\% of the total amount), whereas for the contracting authorities, the top 10\% of the companies spent around 87\% of the total amount (and the top 1\% of the companies is responsible for approximately 60\% of spendings). These are well above {the} standard Pareto's "80-20 rule" \cite{Pareto1971} where {the} top 20\% members of a specific group posses 80\% of the total money amount (or more generally, 20\% of {the} causes are responsible for 80\% of {the} results). Both {the} spendings on and {the} revenues from the public procurement programs are strongly concentrated.

{The statistical accuracy of the power-law fits and the actual closeness of the empirical distributions to the power-laws in the right tail have been tested with the procedure proposed by Preis \textit{et al.} \cite{Preis2011}. Both for the total revenues and the total spendings, we simulate samples with the same number of observations, the same cut-off points and the estimated $\alpha$. The samples are simulated 10,000 times and for each sample, the Kolmogorov-Smirnov test \cite{Stephens1974} is applied to the $cdf$. By doing so, we obtain the critical values and p-values to test whether the distributions of the revenues and spendings are close to being power-law distributed. The test statistics are 0.0007 and 0.0014 for total costs and revenues, respectively. The corresponding p-values are 0.7541 and 0.3820, respectively. Therefore, we cannot reject that total revenues and total costs follow the power-law distribution with the cut-off point at a unit of the corresponding standard deviation. This statistically supports the graphical evidence for the power laws presented in Figs.~\ref{fig2}-\ref{fig3}.}

There are several possible explanations for the observed distributions which might be a subject of the further research. {As an example, t}he distribution of {the} spendings may partially follow from the population of the contracting authorities{,} 48\% of which are municipalities whose population is well {documented} to follow {the} power law distribution \cite{Benguigui2007,Cordoba2008}. Since the population is tightly connected with an economic turnover and thus {the} PP volume, this subpopulation may have a substantial effect on the overall distribution. However, {the} municipalities make roughly {only} 20\% of the total spendings and have only 5 representatives among {the} 100 largest authorities, making it {a rather} weak explanation for the power law tail. Further on, we propose more a general mechanism which might cause emergence of the power law scaling also across other authorities. The distribution of winners is{, however,} a much more interesting result as it emphasizes a massive inequality in distribution of public money that has no straightforward economic justification. The underlying mechanics are likely to be connected with a fact that {the} past won contracts contribute to a chance of winning in a new contract -- either formally as a reference, or through some informal advantage such as an emerging clientelism or corruption ties. Studying this phenomenon on more detailed level is certainly a fruitful area of research. In the rest of this paper, we propose an approach which results in the observed distribution through much simpler means of {the} entropy maximization given reasonable constraints.

Let $M$, $C$ and $Z$ stand for a total amount of money spent on PP, a total number of {the} firms with at least one won contract and a total number of {the} contracting authorities, respectively\footnote{For simplicity, we assume that $M$, $C$ and $Z$ are exogenous. For $M$ and $Z$, this assumption is very reasonable because the total amount of money spent on {the} public procurements as well as the number of the contracting authorities are mainly a political decision. For $C$, this might be an oversimplification since the number of {the} firms that won at least one procurement arises as a solution of some optimization problem. The value of $C$ is a consequence of an economic friction caused by a limited number of firms and costs of entering the PP market.}. Let us further define $C=\sum_{n=1}^N{c_n}$  and $Z=\sum_{k=1}^K{z_k}$ where $c_n$ is a number of companies with {the} total revenues from {the} public procurements of some specific level $n$ and $z_k$ is a number of authorities with {the} total spendings on {the} public procurements of some specific level $k$. Here, $n=1,\ldots,N$ and $k=1,\ldots,K$ are discrete levels of obtained or spent money, respectively. We denote $m_{c_n}$ as a specific amount of revenue of a supplier in $c_n$ so that $\sum_{n=1}^N{c_nm_{c_n}}=M$ and in a similar way{,} $m_{z_k}$ is a specific amount spent on {the} public procurements for an authority in $z_k$ so that $\sum_{k=1}^K{z_km_{z_k}}=M$. Now, we can define a probability that a firm has a total revenue $m_{c_n}$ as $p(c_n)=c_n/C$ and a probability that an authority has spent a total of $m_{z_k}$ on {the} public procurements as $p(z_k)=z_k/Z$. Obviously, it holds that $\sum_{n=1}^N{p(c_n)}=\sum_{k=1}^K{p(z_k)}=1$, $p(c_n)\ge 0$ and $p(z_k)\ge 0$ for all $k$ and $n$, which is needed for a probability measure.

Such a framework {provides} enough information to analyze {the} probability distributions maximizing {the} entropy of the system. For simplicity, we choose {the} supplier side of the transaction so that we work with variables $M$, $C$, $c_n$, $m_{c_n}$, $p(c_n)$ and $N$. Using the definition of $p(c_n)$, we can rewrite the restrictions on $c_n$ as restrictions on probabilities, i.e. $\sum_{n=1}^N{p(c_n)}=1$ and $\sum_{n=1}^N{p(c_n)m_{c_n}}=M/C$. In economics, it is usually assumed that the system is in equilibrium or very close to it. In physics, such a system can be analyzed with a use of entropy and the entropy-maximizing (the most probable) configuration of the distribution is found through a solution of a Lagrangian given constraints{, which is parallel to the maximum likelihood approach used in economics \cite{Aoyama2010}}. {Important aspect of the systems' description is extensivity, i.e. whether the parts of the system are independent (or only weakly dependent/interacting) or strongly dependent/interacting. Therefore, we consider both the extensive and non-extensive systems to see whether the optimization under the given constraints leads to the distributions observed in the public procurement market. For the extensive systems, we utilize the Shannon's entropy and for the non-extensive systems, i.e. systems with strongly interacting particles, we utilize the Tsallis' entropy.} 

Starting with the {extensive systems}, we maximize the Shannon's entropy {\cite{Shannon1948}} $S=-\sum_{n=1}^N{p(c_n)\log(p(c_n))}$ with constraints $\sum_{n=1}^N{p(c_n)}=1$ and $\sum_{n=1}^N{p(c_n)m_{c_n}}=M/C$ yielding the Lagrangian $L_1$:

\begin{multline}
\label{L1}
L_1=-\sum_{n=1}^N{p(c_n)\log(p(c_n))}-\lambda_1\left(\sum_{n=1}^N{p(c_n)}-1\right)-\\
\kappa_1\left(\sum_{n=1}^N{p(c_n)m_{c_n}}-\frac{M}{C}\right)
\end{multline}

{The m}aximization of $L_1$ with respect to $p(c_n)$ gives $p(c_n)=e^{-\kappa_1 c_n+\lambda_1-1}$, where $\kappa_1$ and $\lambda_1$ are Lagrange multipliers respecting the restrictions {or in economic terms, the sensitivities with respect to the given constraints. Interestingly, $\kappa_1$ characterizes the sensitivity to the changes in the average revenue earned by the firms}. Therefore, the maximization of {the} entropy {of the extensive system} yields the Maxwell-Boltzmann (exponential) distribution with an inverse temperature given as the Lagrange multiplier $\kappa_1$.

Considering {the non-extensive systems}, we maximize the Tsallis' entropy {\cite{Havrda1967,Tsallis1988}} $S_q=(1-\sum_{n=1}^N{p(c_n)^q})/(q-1)$, where $q$ is an entropic index, with the same constrains{,} and the Lagrangian $L_2$ is given as:

\begin{multline}
L_2=\frac{1-\sum_{n=1}^N{p(c_n)^q}}{q-1}-\lambda_2\left(\sum_{n=1}^N{p(c_n)}-1\right)-\\
\kappa_2\left(\sum_{n=1}^N{p(c_n)m_{c_n}}-\frac{M}{C}\right)
\end{multline}

Here, the maximization of $L_2$ with respect to $p(c_n)$ yields $p(c_n)=\left(\frac{q-1}{q}\lambda_2+\kappa_2 c_n\right)^{-\frac{1}{q-1}}$, where again $\kappa_2$ and $\lambda_2$ are the Lagrange multipliers respecting the restrictions. Therefore, the maximization of {the} entropy in the {non-extensive system} yields the {Pareto (power law)} distribution. This is indeed what we have observed for {the} total revenues of the participating firms and this process can thus be well described as emerging from the {non-extensive system with strongly interacting particles}. In the same way, this can be shown for the contracting authorities and the distribution of their total spendings where we also found the {power law} distribution. Note that $q>0$ is a measure of a non-extensiveness and the further $q$ is from one, the more non-extensive the system is. For $q\rightarrow 1$, the Shannon's entropy is recovered {(an extensive system)} and the resulting distribution is also exponential as for $L_1$ in Eq. \ref{L1}. 

In a very similar way, we can approach the number of bidders distribution problem. We have $W$ tenders and $K$ distinct values of competing firms for a single tender. Let $w_k$ be a number of tenders with $b_k$ bidders where $k=1,\ldots,K$ so that $\sum_{k=1}^K{w_k}=W$. A probability that there are $b_k$ bidders for {the} tender is $p_k=w_k/W$ and it obviously holds that $\sum_{k=1}^K{p_k}=1$ and $p_k\ge 0$ for all $k$. Adding a constraint on the average value of bidders on a single tender $\sum_{k=1}^K{p_kb_k}=F/W$, which follows from a restriction on a total number of bidding firms for all contracts $F$ defined as $\sum_{k=1}^K{w_kb_k}=F$, we can again construct the Lagrangian form $L_3$ maximizing the Shannon's entropy 

\begin{multline}
L_3=-\sum_{n=1}^N{p_k\log p_k}-\lambda_3\left(\sum_{k=1}^K{p_k}-1\right)-\\
\kappa_3\left(\sum_{k=1}^K{p_kb_k}-\frac{F}{W}\right),
\end{multline}

yielding a probability distribution function $p_k=e^{-\kappa_3b_k-1+\lambda_3}$, i.e. the Maxwell-Boltzmann distribution, which is indeed observed for the number for bidders for our dataset. {The n}umber of bidding firms for a contract thus seems to be generated from {the extensive system with no or only weak interactions between participants}.

To further expand the analogy between {the extensive and non-extensive} systems in physics and our specific socio-economic problem, consider the particles in the physical systems as market participants listing as well as competing for {the} public procurements. As we can hardly describe the behavior of each individual firm, it suffices to analyze only the aggregate behavior. Market participants (particles) can either act independently {(or weakly interact)} or strongly interact, which is a parallel to {the} physical {extensive and non-extensive systems}, respectively. The analogy can be also expanded to potential frictions in the market (collisions of particles) which further increase the entropy and drive the system away from equilibrium. Therefore, the market situations when the market participants do not (or only weakly) cooperate and/or there are no barriers to enter the market can be taken as the {extensive} system, which, as we have shown, {would lead} to the Maxwell-Boltzmann distribution of {the} revenues and spendings. Reversely, the market situation when the market participants cooperate and/or there are barriers and frictions in the market can be taken as the {non-extensive} system, which we have shown to yield the Pareto distribution for {the} spendings and revenues. This even leads us to a policy implication -- the closer the distribution of {the} revenues or spendings is to the Maxwell-Boltzmann distribution, the more transparent and competitive the whole system is, and reversely, the closer the distribution of {the} revenues and spendings is to the Pareto distribution, the less transparent and competitive the whole system is. {Moreover, getting closer to the extensive system characteristics requires the procurement process to be more transparent, with less cooperation between interacting participants (agents) and less market frictions (barriers).} Removing or at least suppressing these inefficiencies shall lead to the Maxwell-Boltzmann distribution of {the} total revenues and spendings, which is characteristic for the {extensive} systems. Even though such a policy advice seems obvious, it is quite strong as it is based on a well defined statistical analysis.

Taking the results for {the} number of bidders into consideration as well, we argue that this process is {similar to an extensive system in physics}. From an economic standpoint, it seems that companies competing for the procurements do not cooperate on the level of bidding, which results in the Maxwell-Boltzmann distribution for the number of bidders. Therefore, the forces driving the total revenues from the {"ideal situation"} seem to be more caused by the customer--supplier cooperation than the supplier--supplier cooperation. This is indeed rather disturbing result which indicates {potential} corruption in the procurement process in the Czech Republic, which is of course illegal and should be a warning for {the} authorities. {However, there might be different causes of such results. The fact that the Pareto distribution is found for both the revenues and spendings indicates that the distribution form might be inherited from the contracting authorities to the suppliers. As the contracting authorities are usually politically influenced and interconnected, it seems obvious that there are strong interactions between them leading to the power-law distribution. The policy implications would thus be much stronger if we found the Maxwell-Boltzmann distribution for the spendings and the Pareto distribution for the revenues. Connected to this, the volume of specific procurements might be complicating the interpretation as well.}

To conclude, we have shown that {the} public procurement market can be analyzed with {the} tools standardly used in {the statistical} physics and not only do these tools give technically interesting results such as the exponential and power laws but they can even lead us to {the} specific policy implications {(even though these should be taken with caution)}. These basic results can be used for further analysis and modeling of processes connected to {the} public procurements. {Note that} the analysis presented here is far from being complete and there are other issues, which should be analyzed in future -- relationship between {the} number of bidders and {the} final price of a contract, between {the} number of won contracts and {the} final price of a contract, {the} concentration of firms with respect to {the} specific contracting authority, and others. Indeed, depending on {the} data availability, it would be interesting whether {the} properties presented here are found for other countries as well.\\

\acknowledgements The authors acknowledge financial support of the Grant Agency of the Czech Republic (grant numbers P402/11/0948 and 402/09/0965), the Technological Agency of the Czech Republic (grant number TD010133), the Grant Agency of the Charles University (grant numbers 118310) and project {SVV 265 504}.

\bibliography{procurement}
\bibliographystyle{eplbib}

\onecolumn

{
\begin{figure}[htbp]
\center
\begin{tabular}{cc}
\includegraphics[width=3.3in]{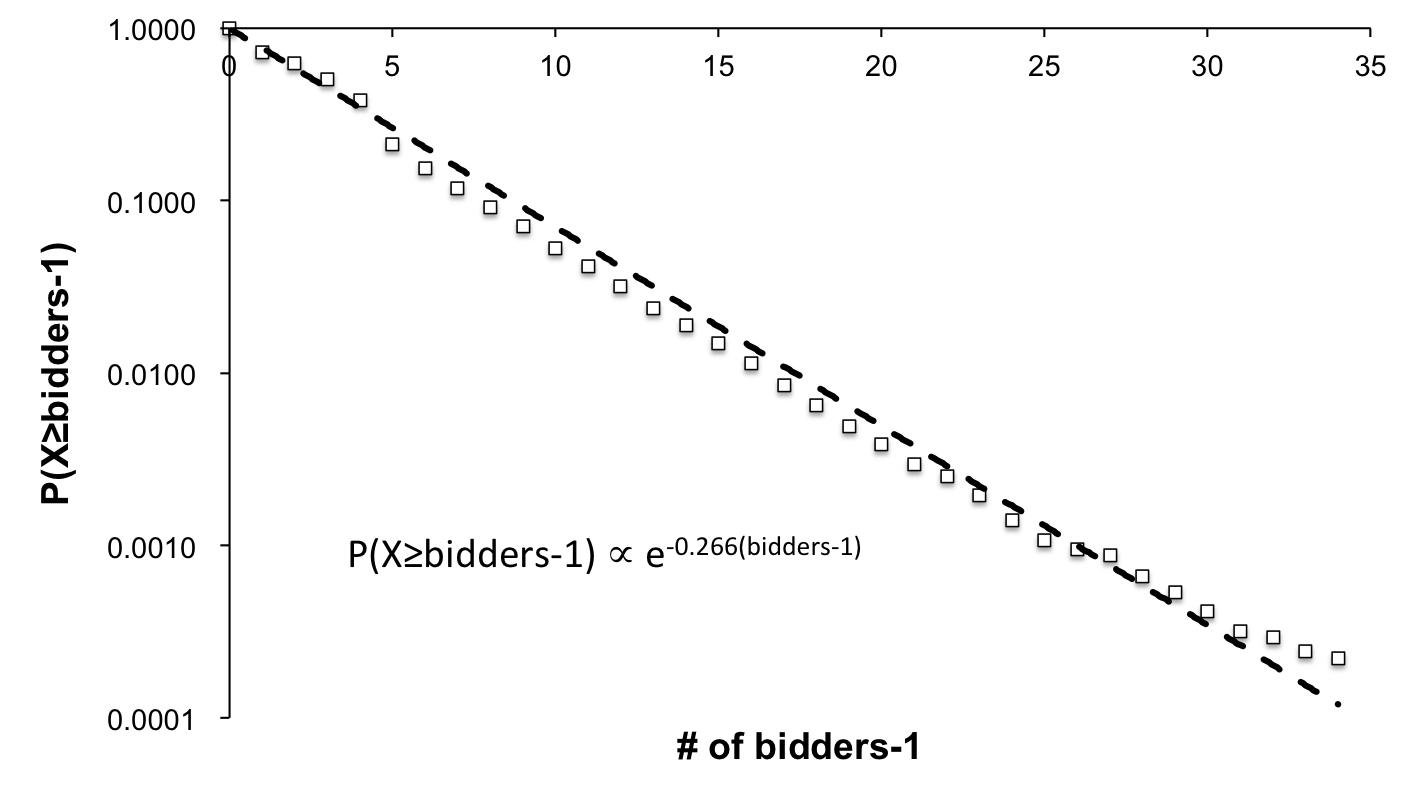}&\includegraphics[width=3.3in]{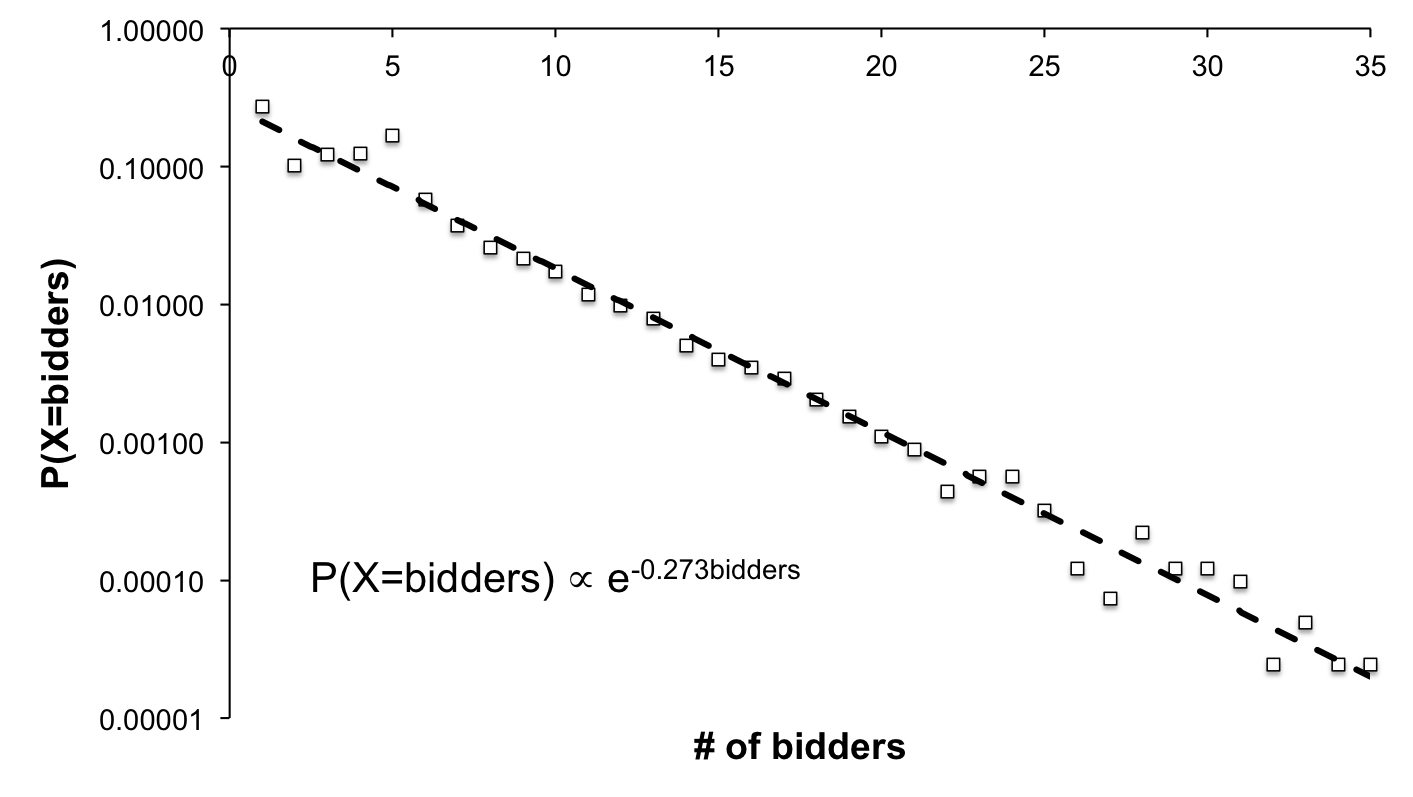}\\
\end{tabular}
\caption{\textit{Distribution function of number of bidders.} Obvious exponential scaling of both $cdf$ (left) and $pdf$ (right) is shown with $\beta \approx 0.27$. As $cdf$ has to equal 1 for the number of bidders equal to 1, the fit is based on a fixed intercept. \label{fig1}}
\end{figure}
}

{
\begin{figure}[htbp]
\center
\begin{tabular}{cc}
\includegraphics[width=3.3in]{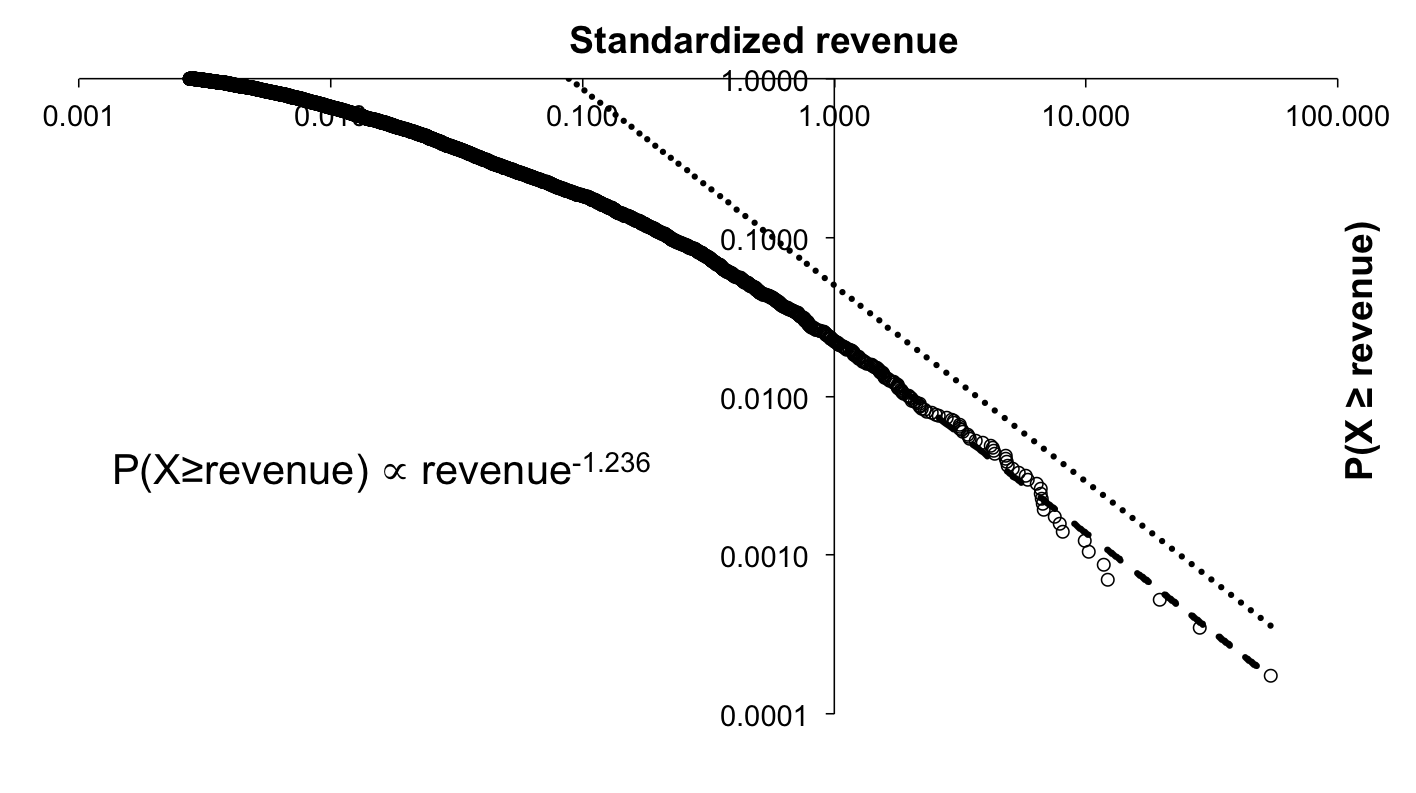}&\includegraphics[width=3.3in]{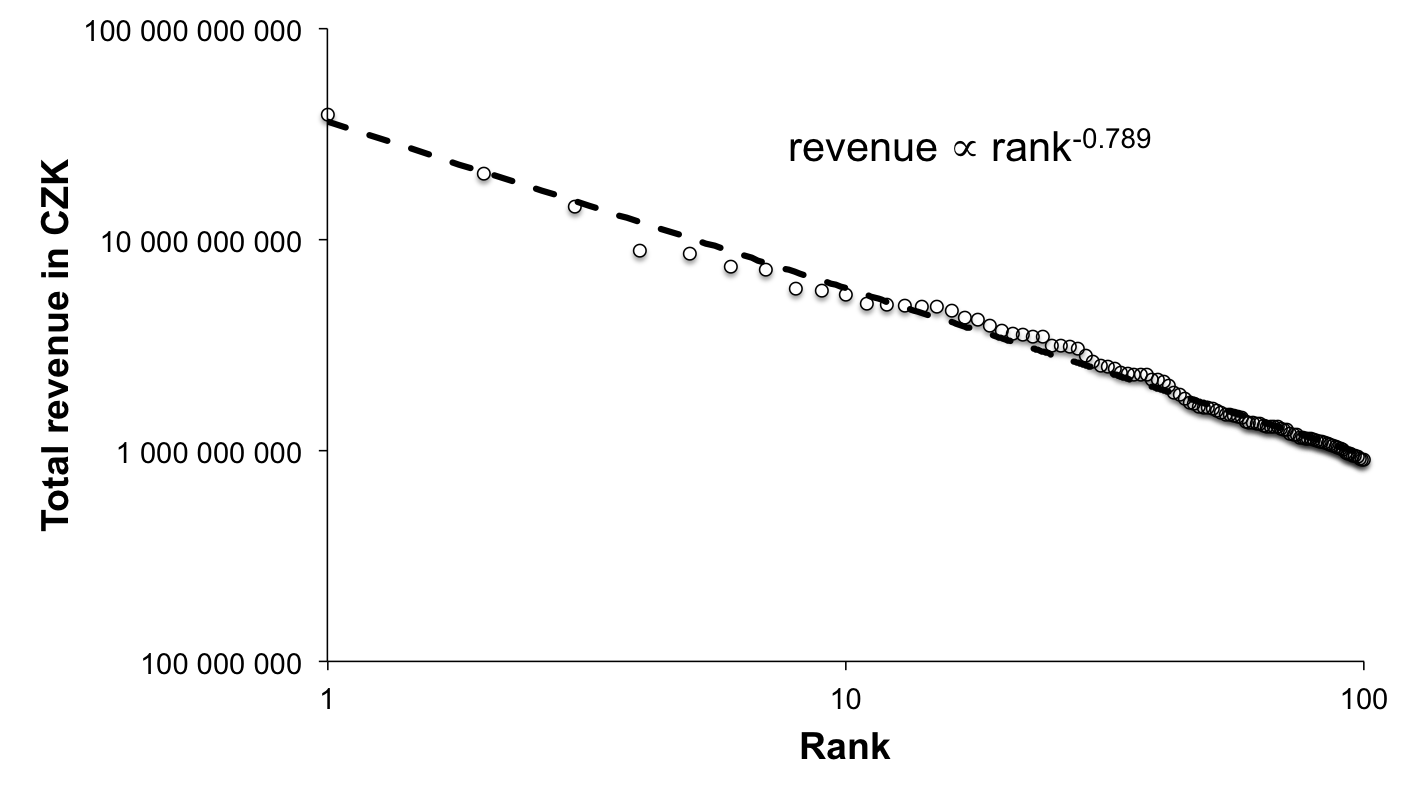}\\
\end{tabular}
\caption{\textit{Distribution function for total supplier revenues.} Revenues are standardized so that they are shown in a number of standard deviations. The power law fit is based on the standardized revenues above a single standard deviation. The power law exponent of $\alpha=1.236$ fits almost perfectly for the right tail (left). The parameter is supported in the Zipf's law plot (right) with $\gamma=0.789$.\label{fig2}}
\end{figure}
}

{
\begin{figure}[htbp]
\center
\begin{tabular}{cc}
\includegraphics[width=3.3in]{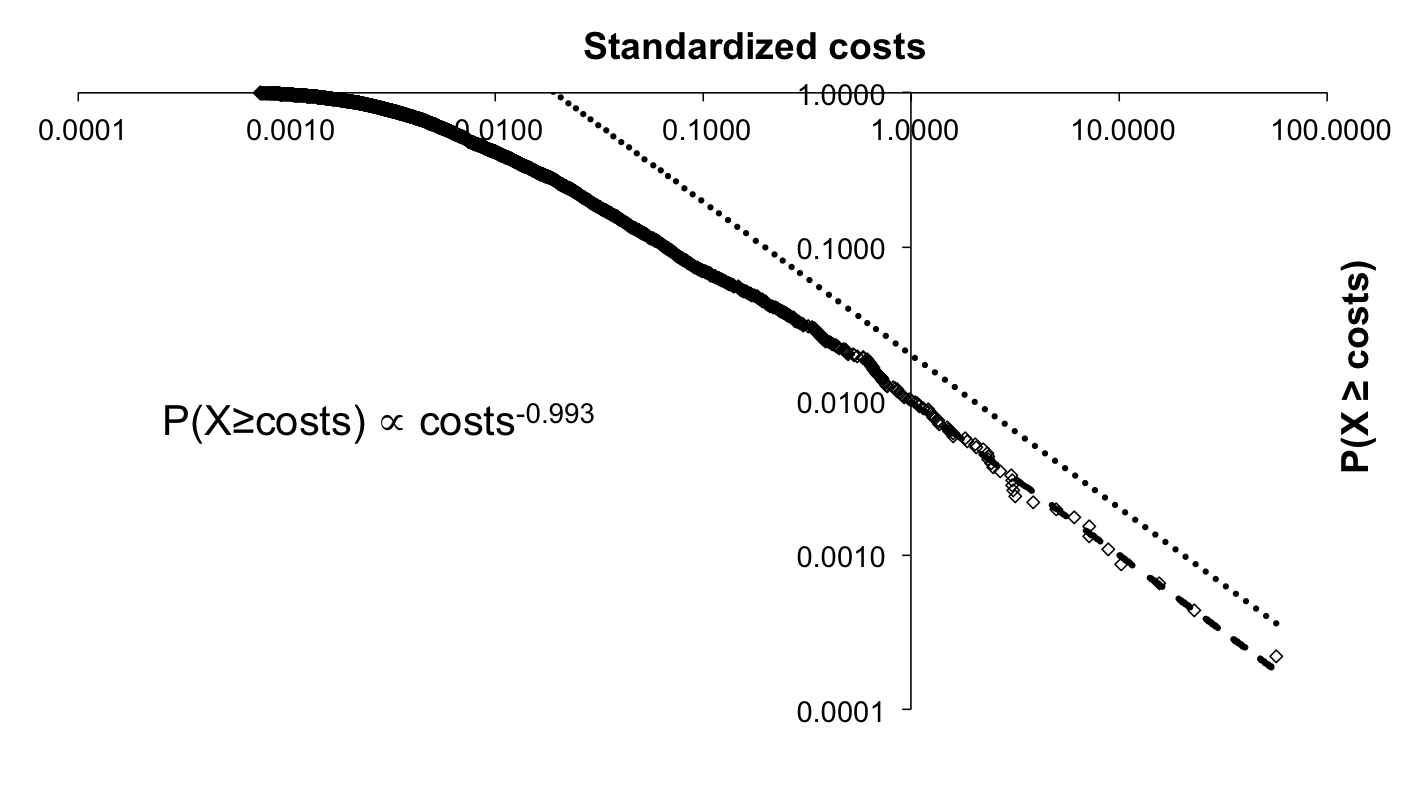}&\includegraphics[width=3.3in]{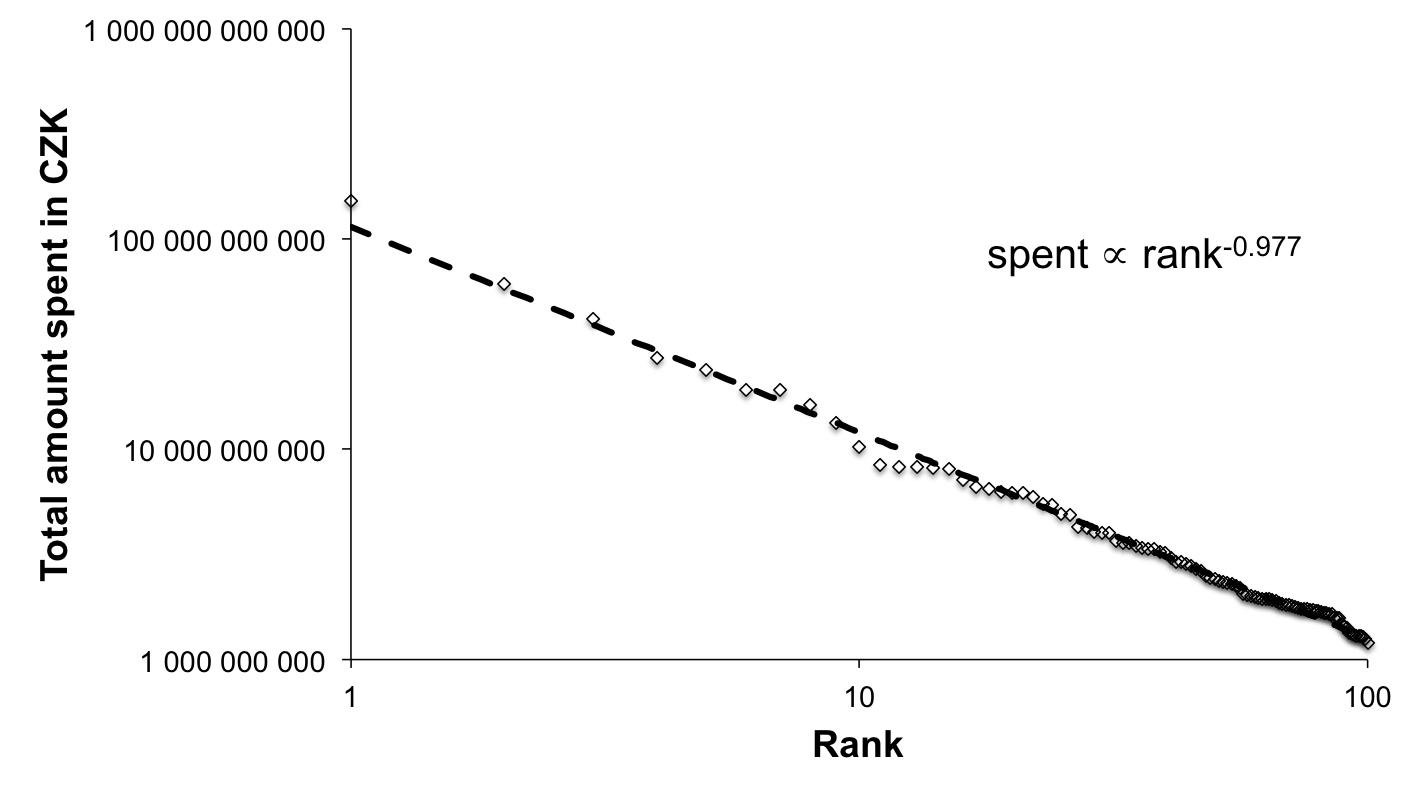}\\
\end{tabular}
\caption{\textit{Distribution function for total spendings of contracting authorities.} Spendings are standardized so that they are shown in a number of standard deviations. The power law fit is based on the standardized spendings above a single standard deviation in the same way as for the revenues. The power law exponent of $\alpha=0.993$ fits almost perfectly for the right tail and holds well even for the lower values of the spendings (left). The parameter is supported in the Zipf's law plot (right) with $\gamma=0.977$. \label{fig3}}
\end{figure}
}

\end{document}